\documentstyle[editedvolume,numreferences,epsf]{crckapb} 

\newcommand{\ket}[1]{{\left | {#1} \right\rangle}}
\newcommand{\bra}[1]{{\left\langle {#1} \right |}}

\begin{opening}
\title{Noise in the Single Electron Transistor and its back action
during measurement}

\author{G. Johansson}
\institute{Institut f\"ur Theoretische Festk\"orperphysik\\
           Universit\"at Karlsruhe, D-76128 Karlsruhe, Germany}
\author{P. Delsing, K. Bladh, D. Gunnarsson, T. Duty, A. K\"ack, G. Wendin}
\institute{Microtechnology Center at Chalmers MC2,\\ 
Department of Microelectronics and Nanoscience,\\
Chalmers University of Technology and G\"oteborg University, S-41296}
\author{A. Aassime}
\institute{Service de Physique de l'Etat Condens\'e, CEA-Saclay, F-91191
Gif-sur-Yvette,France}

\end{opening}

\runningtitle{SET BACKACTION}

\begin{document}


\begin{abstract}
Single electron transistors (SETs) are very sensitive electrometers and they 
can be used in a range of applications. In this paper we give an 
introduction to the SET and present a full quantum mechanical 
calculation of how noise is generated in the 
SET over the full frequency range, including a new formula for the 
quantum current noise. The calculation agrees well 
with the shot noise result in the low frequency limit, and with the 
Nyquist noise in the high frequency limit. We discuss how the SET and in 
particular the radio-frequency SET can be used to read out charge based 
qubits such as the single Cooper pair box. 
We also discuss the backaction which the SET will have on the 
qubit. The back action is determined by the spectral power of voltage 
fluctuations on the SET island. We will mainly treat the normal state 
SET but many of the results are also valid for superconducting SETs.
\end{abstract}



\section{Introduction}
The single-electron transistor (SET) is known as a highly
sensitive electrometer \cite{Fulton,Likharev} based on the Coulomb
blockade \cite{Averin&Likharev,Grabert}. Electrons tunnel one by one
through two small-capacitance tunnel junctions, with capacitances
$C_L$ and $C_R$. The gate, with capacitance $C_g$, is used to modulate 
the generated current, and the island is also capacitively coupled
to the system to be measured through $C_c$. 
The charging energy $E_C= e^2 / 2 C_{\Sigma}$, $C_{\Sigma}=C_L+C_R+C_g+C_c$,
associated with a single electron prevents sequential tunneling
through the island at voltages below a threshold $V_t$, which
can be controlled by applying a voltage $V_g$ to the gate. This
Coulomb blockade is effective at temperatures $T<E_C / k_B$
and for junction resistances larger than the resistance quantum
$R_K=h/e^2\approx25.8 \mathrm{k\Omega}$.

SETs can be made using several different technologies, for example 
they can be fabricated from metallic (often aluminum based) systems, 
from quantum dots in GaAs and silicon, or from carbon nanotubes. 

Often the SETs are operated at low temperatures, however room temperature 
operation has been demonstrated in several cases\cite{Matsumoto,Pashkin,Kim}.

Both the ultimate sensitivity and the back action of SET during 
measurement is determined by the noise in the SET. By understanding 
the noise in the SET we can optimize the use of the SET for each 
application.

\subsection{RF-SET}
The conventional SET, based on measuring either the current or voltage
across the transistor, has suffered from the relatively large output
resistance R of the transistor. For the typical resistance values of
$100\, \mathrm{k\Omega}$ and cable capacitance of $C \sim 1 \mathrm{nF}$, the
corresponding RC time limits the bandwidth to a few kHz. 

With the invention of the radio-frequency SET (RF-SET)\cite{SchoelkopfSc}
the SET was made fast and very sensitive. 
By connecting the SET very close to a cold amplifier the upper 
frequency limit was improved to about 1 MHz\cite{Pettersson,Pohlen}.
With the invention of the RF-SET\cite{SchoelkopfSc}
frequencies above 100 MHz could be reached.

The operation principle of the RF-SET is similar to that of a radar.
A weak RF-signal (the carrier) is launched via a
directional coupler and a bias tee, towards the tank circuit in which 
the SET is embedded. The reflected signal depends critically on the 
dissipation in the tankcircuit, and thus in the SET. 
This means that the carrier is modulated by the gate signal.
The reflected signal is
amplified by a cold amplifier and a number warm amplifiers. The
reflected signal can be analyzed either in the
frequency domain or in the time domain. Typical carrier 
frequencies are in the range of 0.3-2\,GHz.

Charge sensitivities of
$1.2\cdot10^{-5}$ and $3.2\cdot10^{-6}$ $e/\sqrt{\mathrm{Hz}}$ have been
reached for signal frequencies of 100 and 2 MHz,
respectively\cite{SchoelkopfSc,AassimeAPL}. The RF-SET can be used as
a readout device in applications from very sensitive charge meters and
current standards\cite{Grabert} in which electrons are counted or pumped
one by one, to read out of quantum bits\cite{AassimePRL,Lea,KanePRB,MakhlinRMP}
or to work as photon detectors \cite{SchoelkopfIEEE}.

\section{SET - orthodox theory}
\label{OrthodoxTheory}
\begin{figure}
\centerline{\epsffile{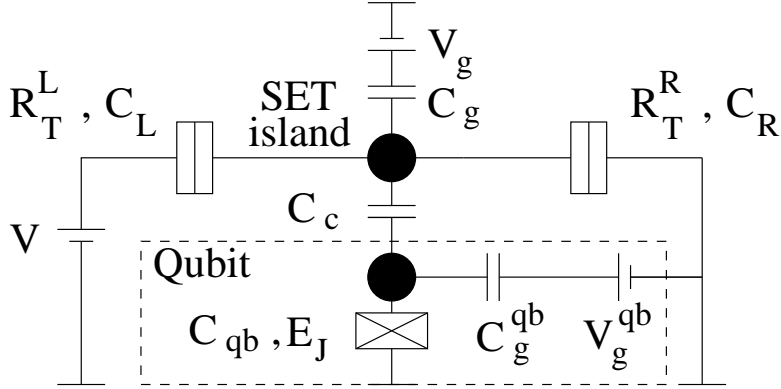}}%
\caption{Schematic figure of the SET capacitively coupled 
to a Single Cooper-pair Box (SCB).}
\label{qnoise_SET_fig}
\end{figure}

Consider a small metallic SET island coupled via low transparency
tunnel barriers to two external leads, and coupled capacitively to 
an object to be measured. In Fig.~\ref{qnoise_SET_fig} we have used the
example of a Single Cooper-pair Box (SCB)\cite{BouchiatPS} which may serve as 
a qubit in a quantum computer. The SCB is controlled by a voltage source 
$V^{qb}_g$.

Following orthodox SET theory\cite{Likharev,Averin&Likharev} we use
the integer number of electrons on the SET island ($N$) as
a basis for describing its dynamics. This is motivated by
the low transparency junctions.
The charge on the island is the sum of the electrons, the
background charge, and the charge induced by the the voltages on the
three island capacitances. Electrostatics gives for the Coulomb charging
energy $E_{ch}$ and island potential $V_I$:
\begin{eqnarray}
\label{electrostatics}
E_{ch}(N)&=&E_C (N-n_x)^2,\ \ E_C=\frac{e^2}{2C_\Sigma}, \nonumber \\
V_I(N)&=&\alpha_L V-\frac{e(N-n_x)}{C_\Sigma},\ \
\alpha_L=\frac{C_R}{C_\Sigma},\ \
\alpha_R=\frac{C_L}{C_\Sigma},
\end{eqnarray}
where $n_x e=V_g^{qb} C_g^{qb}$ is the charge induced on the
gate capacitance $C_g^{qb}$ by the external voltage
source $V_g^{qb}$.
$\alpha_{L/R}$ describes the capacitive voltage division over the
$L/R$ junction, were we have neglected the small gate and coupling
capacitances $\{C_c,C_g\} \ll \{C_L, C_R\}$. 

The orthodox theory neglects mainly three things: the effect of the 
electromagnetic environment of the SET \cite{IngoldNazarov}, 
higher order tunneling processes, i.e. cotunneling \cite{Geerligs},
and that for high frequency dynamics the transition rates are
frequency dependent. 
As long as the electromagnetic environment has a low impedance compared 
to the quantum resistance $R_{Q}=h/4e^2 \approx 6.45 \mathrm{k\Omega}$ 
(which is often the case for SETs) the corrections due to the environment 
are small. The same is true for the cotunneling 
corrections as long as the resistance of the SET is large compared 
to $R_{Q}$.
To take into account the frequency dependence of the transition rates,
including the energy exchange with the measured system, is the main 
objective of the quantum theory presented in section~\ref{QuantumTheory}.

\subsection{Transition rates}
The dynamics of the SET consists of stochastic transitions 
between the charge states, i.e. by electrons randomly tunneling on and 
off the island. In orthodox SET theory the rates for the different
transitions are given by the Golden Rule rates for tunneling through
the left and right tunnel junctions
\begin{equation}
\Gamma^{L/R}(E_{if})=\frac{R_K}{R^{L/R}_T}
\frac{E_{if}}{h}\frac{1}{1-e^{-E_{if}/k_B T}},
\label{GoldenRates}
\end{equation}
where $E_{if}$ is the energy difference between the initial and final state
and the Bose function appears from the convolution of the two Fermi functions
for the filled initial states and the empty final states in the leads.
$E_{if}$ has two terms; one is the change in charging energy, and the other
is the work done by the voltage bias.

To be specific we now consider the electrons to gain energy from the
voltage source by tunneling from left to right. Then the rates
$\Gamma_{n\pm}^{L/R}$ for transitions from charge state
$n$ to $n\pm1$ across the left/right junction are 
\begin{equation}
\begin{array}{cc}
\Gamma_{n+}^{L}=\Gamma^{L}(E_{n(n+1)}+eV_L), &
\Gamma_{n+}^{R}=\Gamma^{R}(E_{n(n+1)}-eV_R), \\
\Gamma_{n-}^{L}=\Gamma^{L}(E_{n(n-1)}-eV_L), &
\Gamma_{n-}^{R}=\Gamma^{R}(E_{n(n-1)}+eV_R),
\end{array}
\end{equation}
where $E_{mn}=E_{ch}(m)-E_{ch}(n)$ and $V_{L/R}=\alpha_{L/R}V$ is the
voltage drop over the $L/R$ junction. We denote the sum of rates taking
the SET island from $n$ to $n\pm1$ with 
$\Sigma_{n\pm}=\Gamma_{n\pm}^{L}+\Gamma_{n\pm}^{R}$.

\subsection{Master equation and steady-state}
\begin{figure}
\epsffile{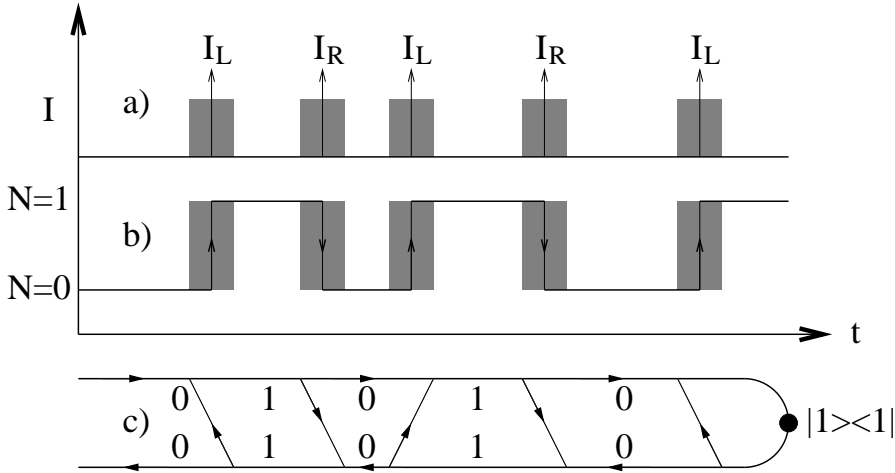}%
\caption{In orthodox theory the transitions between charge states in the SET
are considered to be instantaneous.
For a specific realisation (measurement) this results in: 
a) The currents through the left/right junctions $I_L/I_R$ consist of a 
train of delta-peaks.
b) The charge on the island $N(t)$ is a telegraph signal.
In order to calculate the transition rates one
uses the Golden Rule approximation, which includes integration over a
timescale of $h/E_{if}$, where $E_{if}$ is
the energy gain in the transition. This is indicated as grey areas in
figures a) and b).
c) In the real-time Keldysh formalism this timescale of tunneling reappears
as a timescale on which the density matrix is off-diagonal.}
\label{NofTfig}
\end{figure}

During a specific measurement, i.e. a realisation of the stochastic process,
the number of (extra) electrons on the SET island as a
function of time is piecewise constant, with steplike changes due to
instantaneous tunnel events (see Fig.~\ref{NofTfig}a-b).
Mathematically one may describe the dynamics of this
process averaged over a large number of measurements.
Introducing the probability $P_n(t)$ of finding the SET in
charge state $n$ at time $t$ we may write down a master equation
\begin{equation}
\label{master_equation}
\partial_t \bar{P}(t) =
\hat{\Sigma}\cdot\bar{P}(t),
\end{equation}
where $\bar{P} =[\dots P_{-1}(t)\ P_0(t)\ P_1(t) \dots]^T$ is
a column vector and $\hat{\Sigma}$ is the tridiagonal transition rate matrix
with off-diagonal elements
$(\hat{\Sigma})_{(n\pm1)n}=\Sigma_{n\pm}$, and to ensure probability
conservation $(\hat{\Sigma})_{nn}=-(\Sigma_{n+}+\Sigma_{n-})$.
The solution to Eq.~(\ref{master_equation}) may be written
$\bar{P}(t)= \hat{\Pi}(t)\cdot\bar{P}(0)$,
introducing the time-evolution operator (matrix) 
$\hat{\Pi}(t)= e^{\hat{\Sigma}t}$.
To compute the matrix exponent one needs to diagonalize $\hat{\Sigma}$.
The steady-state corresponds to the eigenvector of $\hat{\Sigma}$
with eigenvalue zero, and the other eigenvalues determine rates
for exponential decay of the corresponding eigenvector.

%
%
%
%
\begin{figure}
\centerline{\epsfxsize=12cm\epsffile{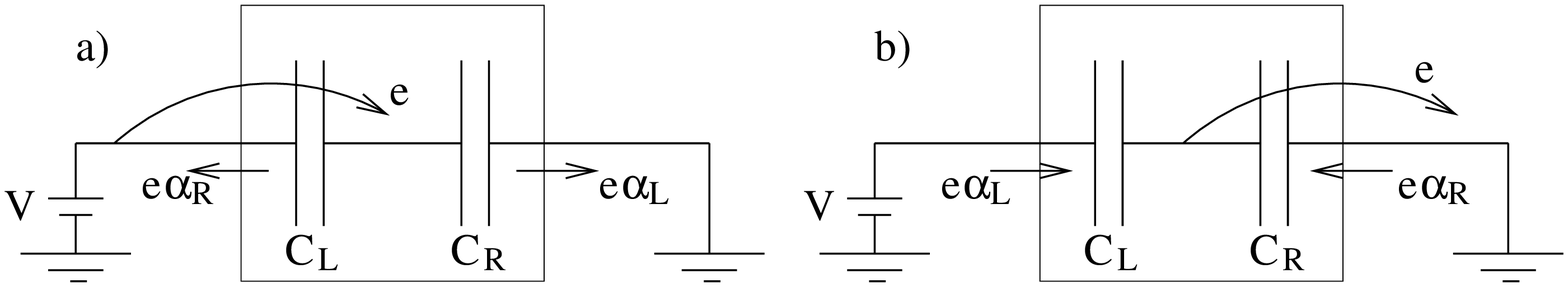}}%
\caption{When an electron tunnels from/to the leads to/from
the island the charges on the capacitances will redistribute.
The charge transport across the boundary of
a region containing the SET island plus its nearby capacitances
is shown for 
a) an electron tunneling onto the island from the left lead, and
b) an electron tunneling from the island into the right lead.
$e \alpha_{L/R}$ represents a displacement charge.}
\label{DispCurrFig}
\end{figure}

\subsection{Charge, Voltage and Current}
In order to calculate physical quantities we introduce operators
for the number of (excess) electrons on the 
island ($\hat{N}$) and the tunnel currents across the left ($\hat{I}_L$)
and right ($\hat{I}_R$) junctions,
\begin{equation}
(\hat{N})_{nn}=n,\ \
(\hat{I}_L)_{(n\pm1)n}=\pm e \Gamma^L_{n\pm}, \ \
(\hat{I}_R)_{(n\pm1)n}=\mp e \Gamma^R_{n\pm},
\end{equation}
noting only the non-zero elements, i.e. $\hat{N}$ is diagonal, 
and $\hat{I}_{L/R}$ are both tridiagonal with zeros on the main diagonal.
The steady-state expectation value $O^{st}$ of an operator $\hat{O}$ 
is given by
\begin{eqnarray}
O^{st}&=&\langle \hat{O} \rangle_{st}\equiv
[1 \dots 1]\cdot \hat{O} \cdot \bar{P}^{st} ,\ \
\bar{P}^{st}=\hat{\Pi}(\infty)\cdot \bar{P}(0)
\Rightarrow \\
N^{st}&=&\sum_n nP_n^{st},
I_L^{st}=e \sum_n P_n^{st} (\Gamma^L_{n+}-\Gamma^L_{n-}),
I_R^{st}=e \sum_n P_n^{st} (\Gamma^R_{n-}-\Gamma^R_{n+}), \nonumber
\end{eqnarray}
which defines the meaning of the brackets $\langle\ \rangle_{st}$.
One gets the voltage of the SET island by multiplying the number
of extra electrons $N$ by $e/C_\Sigma$, and adding the gate bias dependent
constant according to Eq.~(\ref{electrostatics}).
One should also note that each tunnel event in the SET is
followed by a fast redistribution of charge on the nearby
capacitances, see Fig.~\ref{DispCurrFig}. Thus a tunnel event
in one junction will create a displacement current also in the
opposite lead. The operator $\hat{I}$ for the externally measurable
current is given by\cite{Shockley,ButtikerNoiseReview,Korotkov2,Fedichkin,Galperin}
$\hat{I}=\alpha_L \hat{I}_L + \alpha_R \hat{I}_R$,
(see Eq.~(\ref{electrostatics})).
For the steady-state current
we have $I^{st}=I_L^{st}=I_R^{st}$, using the detailed balance\cite{Grim_Stir}
$P^{st}_n \Sigma_{n+}=P^{st}_{n+1} \Sigma_{(n+1)-}$, but for
finite frequency properties the difference between tunnel
currents and the externally measurable current becomes important.

\section{Noise - Fluctuations and Correlations}
The SET dynamics is noisy due to its stochastic nature.
The fluctuations in the current through the SET determine
the measurement time ($t_{m}$) needed to separate the dc currents
corresponding to different charges at the input of the SET.
%
The fluctuations of the charge on the SET island
induce a fluctuating voltage on the capacitance coupling
to the measured system (see Fig.~\ref{qnoise_SET_fig}), 
which may be disturbed. This effect is called
the back-action of the measurement, i.e. the meter acting back
on the measured system.

The fluctuations in the SET can be thought of in terms of two
contributions, one from the shot noise of the sequential tunneling
described by orthodox theory, and one from the quantum fluctuations.
At low frequency, the shot noise will dominate, as described by
orthodox SET theory in section~\ref{OrthodoxNoise} below. 
At high frequencies the SET bias may be neglected and the quantum 
fluctuations, i.e. Nyquist noise, will dominate. 
The Nyquist voltage noise is given by the
impedance of the SET island to ground\cite{CallenWelton}, i.e. from the two
junctions in parallell ($Z_p(\omega)$), while the current noise is given 
by the impedance through the SET, i.e. from the two junctions in series 
($Z_s(\omega)$):
\begin{eqnarray}
\label{NyquistNoise}
S_{VV}(\omega)&=&2\hbar\omega\mathrm{Re}\left\{Z_{p}(\omega)\right\}
\rightarrow \frac{e^2}{C_\Sigma^2}\frac{R_K}{\pi\omega}
\left[\frac{1}{R^L_T}+\frac{1}{R^R_T} \right]+O(\omega^{-2}),\nonumber \\
S_{II}(\omega)&=&2\hbar\omega
\frac{\mathrm{Re}\left\{Z_s(\omega)\right\}}
{|Z_s(\omega)|^2} \rightarrow 2\hbar\omega
\left[\frac{\alpha_L^2}{R^L_T}+\frac{\alpha_R^2}{R^R_T} \right]
+O(1),
\end{eqnarray}
given to leading order in $\omega^{-1}$.

\begin{figure}[t]
\centerline{\epsfxsize=12cm\epsffile{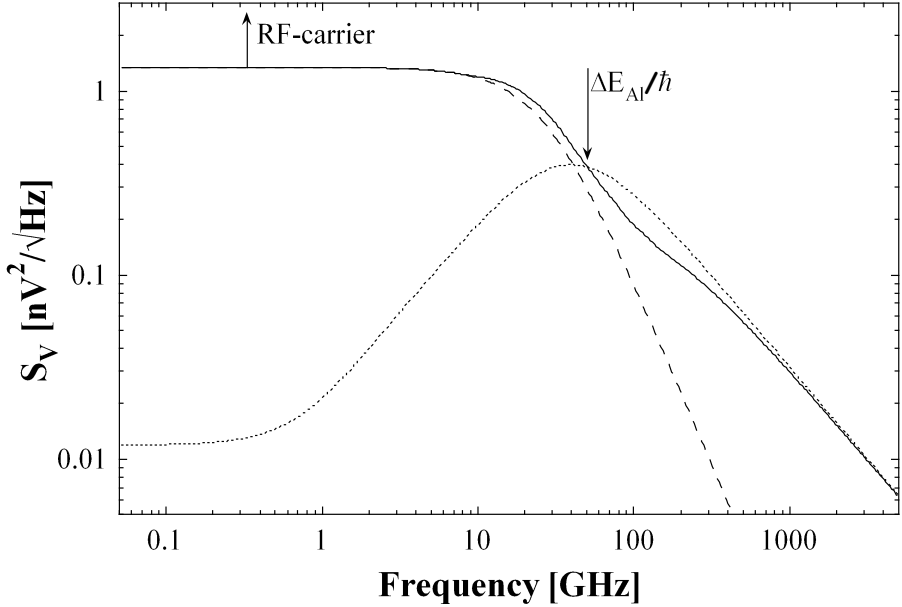}}
\caption{\label{Sv(f)} Comparison of the calculations of the
symmetrized spectral density of voltage fluctuations of the SET island
using the full quantum mechanical calculation (full line), the
classical shot noise (dashed line), and the quantum fluctuations
assuming a linear SET impedance (dotted line).  Parameters for
sample \#1 described in Table\,\ref{Resultstable}, 
are used in all calculations}
\end{figure}

To obtain the spectrum of fluctuations for intermediate frequencies
it is necessary to solve the full quantum problem, which was done
in \cite{JohanssonPRL} for the voltage noise.  The result of this
calculation and the comparison to the shot noise and quantum fluctuation
results in the two limits, are shown in Fig.\,\ref{Sv(f)} for sample \# 1
described in Table\,\ref{Resultstable}.
As can be seen, the result of the full calculation coincides with
the shot noise and the quantum noise in the low and high frequency
limits respectively.

In addition to this spectrum, the RF excitation gives a component at
$f_{RF}$, and the non-linearity of the IV-characteristics gives
an additional component at 3$f_{RF}$. However these frequencies are
much lower than the relevant mixing frequency $\Delta E/\hbar$ for
the SCB-qubit described in section~\ref{qubitsection}.

\subsection{Noise in the SET: Orthodox theory}
\label{OrthodoxNoise}
Since orthodox SET theory gives the correct low-frequency limits for the 
noise, and is also the natural reference for discussing the quantum theory,
we will describe this in some detail.
Together with section~\ref{OrthodoxTheory}
the following discussion is sufficient for calculating the orthodox
theory noise, including an arbitrary number of charge states. 
In section~\ref{Two-Level} we will go through the analytically 
solvable case of two charge states.

We are interested in the fluctuation of charge and current around their
average values. Therefore we define fluctuation operators, 
$\delta \hat{O}=\hat{O}-O^{st} \mathbf{1}$, where $\mathbf{1}$ is the
unit matrix with the same dimension as $\hat{O}$ and $O^{st}$ is the
steady-state expectation value of the operator. We may then write 
the following expressions for the fluctuation correlation function ($\tau>0$)
\begin{equation}
K_{AB}(\tau) = \langle \delta\hat{A} \cdot \Pi(\tau) \cdot
\delta\hat{B} \rangle_{st}, \ \
K_{AB}(-\tau)=K_{BA}(\tau) .
\end{equation}
The master equation (\ref{master_equation}) is dissipative and therefore 
irreversible in time, which explains the need for the special 
negative $\tau$ definition.
Special care has to be taken for the autocorrelation of a current
pulse, i.e. the shot-noise. Since the first current operator instantaneously
changes the state of the system, the second current operator does
not operate on the same state, and thus the shot-noise is not included.
Using some arbitrary representation for the $\delta$-function current pulse
on the tunneling time-scale $E_{if}/\hbar$ one gets
$\int \delta(t-t_0)\delta(t) dt = \delta(t_0)$ which gives
\begin{equation}
K_{I_rI_r}(\tau) \rightarrow K_{I_rI_r}(\tau)+
e\delta(\tau)\sum_n P_n^{st}(\Gamma^r_{n+}+\Gamma^r_{n-}),
\ \ r\in\{\L,R\}.
\end{equation}
%
We will also need the spectral densities of the fluctuations,
i.e. the Fourier transforms of the correlation functions
\begin{equation}
\label{spectral_densities}
S_{AB}(\omega)=\int_{-\infty}^{\infty} e^{-i\omega \tau}
K_{AB}(\tau) d\tau =
\int_{0}^{\infty}
e^{-i\omega \tau} K_{AB}(\tau) + e^{i\omega \tau} K_{BA}(\tau) d\tau.
\end{equation}
By e.g. diagonalizing $\hat{\Sigma}$ one finds that the needed Laplace
transform of $\hat{\Pi}(\tau)$ is given by Eq.~(\ref{Dyson}), just
replacing $\hat{\Sigma}(\omega)\rightarrow\hat{\Sigma}$.

Here we use the unsymmetrized definitions of the correlation functions
in order to separate noise where the tunnel processes absorb the energy
$|\hbar\omega|$ (positive frequencies) from noise where the tunnel
processes emit the energy $|\hbar\omega|$ (negative frequencies), when 
we go to the quantum expressions in the next section.
When this separation is not needed one may use the symmetrized
definition $S_{AB}^{sym}(\omega)=\int_{-\infty}^{\infty} e^{-i\omega\tau} 
(K_{AB}(\tau)+K_{BA}(\tau))d\tau = S_{AB}(\omega)+S_{AB}(-\omega)$.
In the classical calculations $S_{II}(\omega)=S_{II}(-\omega)$; therefore
the symmetric definition only gives a factor of 2 needed to recover
the usual expression for the shot-noise across a single junction
$S_{II}^{sym}(\omega)=2 e I^{st}$\cite{Schottky}.

Apart from using unsymmetrised correlation functions, the formalism
for the orthodox theory presented in this section is equivalent to, 
and inspired by, the work of Korotkov\cite{Korotkov2}.

\subsection{Noise in the SET: Quantum Theory Results}
\label{QuantumTheory}
The master equation (\ref{master_equation}) contains only dissipative
transitions between SET charge states. The fact that the SET might be
in a coherent state like $|N=0\rangle+|N=1\rangle$ was only taken
into account in deriving the tunneling rates. This Golden Rule derivation
includes an integration over the timescale $t_{GR}=h/E_{if}$, and the main
approximation behind Eq. (\ref{master_equation}) is that this timescale is
short compared to the dynamics you describe. For steady-state or
low-frequency properties this is fulfilled since the tunneling
rates include the small tunnel conductance, i.e. 
$1/\Sigma \sim t_{GR} R_K/R_T \ll t_{GR}$.
The shot-noise term had to be handled separately since it
corresponds to correlations on the time-scale $t_{GR}$.
For high frequency dynamics, i.e. $\hbar\omega \sim E_{if}$,
the approximation breaks down and one has to consider the
effects of quantum coherence. 

One way to include quantum coherence is to use the real-time diagrammatic
Keldysh approach described by Schoeller and
Sch\"on\cite{SchoellerSchon}. In the sequential tunneling approximation
the low-frequency results coincide with orthodox theory\cite{Korotkov2}, 
and the diagrams used may also be compared with the time evolution
of the state of the SET island, see Fig.~\ref{NofTfig}c. We will
not further describe the method here, only state the results
in such a way that the finite frequency noise, including arbitarily
many charge states, may be calculated.

\subsubsection{Inelastic transition rates}
When we take into account that the SET can absorb or emit energy
the transition rates are modified.
First of all the tunnel rates become frequency dependent,
since the noise energy $|\hbar\omega|$ should be created or absorbed
in a tunnel event. The rates
\begin{equation}
\Gamma^L_{n\pm}(\omega)=\frac{1}{2}
\Gamma^L(E_{n(n\pm1)}\pm eV_L +\hbar\omega), \ \
\Gamma^R_{n\pm}(\omega)=\frac{1}{2}
\Gamma^R(E_{n(n\pm1)}\mp eV_R +\hbar\omega) ,
\end{equation}
are here defined so that the transition is facilitated by positive
$\hbar\omega$, i.e. the tunnel event absorbs energy for positive
$\hbar\omega$. 
The definition of the $\hat{\Sigma}$-matrix below Eq.~(\ref{master_equation})
is still valid replacing the rates with the sum of positive and negative
frequency rates
\begin{equation}
\Sigma^{L/R}_{n\pm}(\omega)=\Gamma^{L/R}_{n\pm}(\omega)+
\Gamma^{L/R}_{n\pm}(-\omega) , \ \ 
\Sigma_{n\pm}(\omega)=\Sigma^L_{n\pm}(\omega)+\Sigma^R_{n\pm}(-\omega).
\end{equation}
Note that in the zero-frequency limit the orthodox transition rates
are recovered, i.e. $\Sigma_{n\pm}(0)=\Sigma_{n\pm}$.
The time-evolution is evaluated by Laplace-transformation, and
the Laplace transform of the time-evolution operator obeys
the following Dyson type of equation\cite{Dyson}
\begin{equation}
\label{Dyson}
\hat{\Pi}(\omega)=\frac{i}{\omega}
\left[\hat{1}-\frac{i}{\omega}\hat{\Sigma}(\omega)\right]^{-1},
\end{equation}
where $\hat{1}$ is the unit matrix with the same dimension
as the $\hat{\Sigma}$-matrix, determined by the number of relevant
charge states.

\subsubsection{Finite Frequency Noise Expressions}
We now present the noise expressions valid at finite frequency.
The charge noise is given by
\begin{equation}
S_{NN}(\omega)=2Re\left[[1 \dots 1]\cdot \hat{N} 
\cdot \hat{\Pi}(\omega) \cdot \hat{N}_1(\omega) \cdot 
\bar{P}^{st}\right],
\end{equation}
where the charge operator now is tridiagonal and frequency dependent when
it stands in the first position ($\hat{N}_1$),
\begin{equation}
(\hat{N}_1)_{(n\pm 1)n}=\mp i\Gamma_{n\pm}(\omega))/\omega, \ \
(\hat{N}_1)_{nn}=-[(\hat{N}_1)_{(n-1)n}+(\hat{N}_1)_{(n+1)n}] .
\end{equation}
The current noise is given by
\begin{equation}
S_{II}(\omega)=S^{shot}_{II}(\omega)+
[1 \dots 1]\cdot \hat{I}_2(\omega) \cdot 
\left[\hat{\Pi}(\omega)+\hat{\Pi}(-\omega)\right]
\cdot \hat{I}_1(\omega) \cdot \bar{P}^{st} ,
\end{equation}
where the shot-noise contribution is
\begin{equation}
S^{shot}_{II}(\omega)=e^2
\sum_{n,r\in\{L,R\}} 2 P_n^{st} \alpha_r^2 \left[\Gamma^r_{n+}(\omega)+
\Gamma^r_{n-}(\omega)\right] ,
\end{equation}
%
%
%
and the current operators are both tridiagonal matrices
with non-zero entries
\begin{eqnarray}
(\hat{I}_{L1})_{(n\pm1)n}&=&
\pm e\left[\Gamma^L_{n\pm}(0)+\Gamma^L_{n\pm}(\omega)\right],
\nonumber\\
(\hat{I}_{L1})_{nn}&=&
e[\Gamma^L_{n+}(0)-\Gamma^L_{n+}(\omega)]-
e[\Gamma^L_{n-}(0)-\Gamma^L_{n-}(\omega)], \nonumber\\
(\hat{I}_{L2})_{(n\pm1)n}&=&\pm e\Sigma^{L}_{n\pm}(\omega)\ , \ \
(\hat{I}_{L2})_{nn}= 0 .
\end{eqnarray}
The operators for the right junction are constructed by
changing $L \rightarrow R$ and multiplying by $-1$.
The total current operator is still
$\hat{I_{n}}=\alpha_L \hat{I}_{Ln}+\alpha_R \hat{I}_{Rn}$.
%
%
%

We see that also the current and charge operators aquire a frequency
dependence and that the first and second operator have different form,
due to correct ordering along the Keldysh contour.

\subsection{Different Limits of the Quantum Expressions}
\begin{figure}
\centerline{\epsfxsize=8cm\epsffile{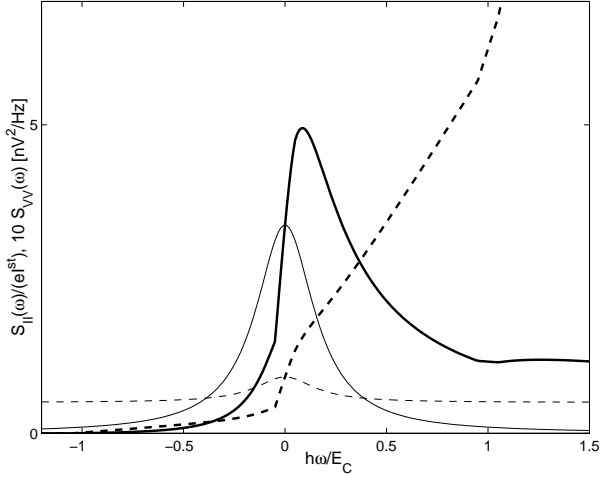}}%
\caption{Calculated $S_{VV}(\omega)$ (bold solid line) and $S_{II}(\omega)$ 
(bold dashed line) for a symmetric SET at zero temperature 
with $R_T^L=R_T^R=R_K$ and 
$E_C=3\, k_B$. The thin lines are the results from orthodox theory.
The SET is biased at $n_x=0.25$, and slightly above the Coulomb 
threshold $eV=1.1 E_C$, with $I^{st}\approx 0.5\mathrm{nA}$. 
Both $S_{VV}(\omega)$ and $S_{II}(\omega)$
go to zero at $\hbar\omega=1.05 E_C$ which correspond to the maximal
extractable energy.}
\label{SI+SVFig}
\end{figure}

An example of $S_{VV}(\omega)$ and $S_{II}(\omega)$
is shown in Fig.~\ref{SI+SVFig}, with the orthodox results as
comparison.
In the limit $\omega\rightarrow0$ the orthodox results are
recovered. The quantum corrections in the low-frequency regime 
are discussed in detail for the two-level system in section
\ref{LowFreqQuantumNoise}.
The noise at large negative frequencies, which correspond to noise
where the SET emits large energies, tend to zero.
In our approximation the voltage noise is analytically zero
when $P_n^{st}\Gamma_{n\pm}(\omega)=0$ for all $n$, i.e. when no
inelastic tunneling events are allowed from the steady-state.
The shot-noise part of the current noise behaves similarly,
and the correction has the same order of magnitude as the
cotunneling shot-noise, which we do not take into account.
Therefore, to the accuracy of our approximation, the current
noise vanishes when the shot-noise vanishes, i.e. when 
$P_n^{st}\Gamma_{n\pm}(\omega)=0$ for all $n$. 

In the high positive frequency limit the spectral noise density of the SET
should be independent of the bias and be dominated by the Nyquist
noise in Eq.~(\ref{NyquistNoise}).
In this limit all rates entering the quantum noise formulas simplify to 
\begin{equation}
\Gamma_{n\pm}^{L/R}(\omega) \rightarrow \Gamma^{L/R} \equiv
\frac{R_K}{2 R_T^{L/R}}\frac{\omega}{2\pi}+O(1)=
\frac{\hbar\omega}{2 e^2 R_T^{L/R}}+O(1),
\end{equation}
where $O(1)$ is a bias-dependent constant.
Using $S_{VV}(\omega)=e^2/C_\Sigma^2\ S_{NN}(\omega)$ we thus
arrive at the following high-frequency quantum noise expressions
%
\begin{equation}
S_{VV}(\omega) \approx \frac{2 e^2}{\omega^2 C_\Sigma^2}
2 (\Gamma^L+\Gamma^R), \ \
S_{II}(\omega) \approx S_{II}^{shot}(\omega)\approx
4 e^2 (\alpha_L^2 \Gamma^L  + \alpha_R^2 \Gamma^R),
\end{equation}
where we used that $\sum_n P_n^{st} =1$. One may easily check that 
this agrees with Eq.~(\ref{NyquistNoise}).

\section{Two level approximation for the SET}
\label{Two-Level}
As an explicit example we now show the analytically solvable case with
low bias and temperature, such that only the two lowest energy
charge states, say $N=0$ and $N=1$, are occupied,
i.e. the only non-zero rates are $\Sigma_{0+}$ and $\Sigma_{1-}$.

\subsection{Orthodox Theory}
The master equation Eq.~(\ref{master_equation}) then simplifies to
\begin{equation}
\label{master_equation_two_level}
\partial_t \left(\begin{array}{c}
P_0(t) \\
P_1(t) \end{array} \right) =
\left(\begin{array}{cc}
-\Sigma_{0+} & \Sigma_{1-} \\
\Sigma_{0+} & -\Sigma_{1-} \end{array}\right)
\left(\begin{array}{c}
P_0(t) \\
P_1(t) \end{array} \right).
\end{equation}
with the solution given by the time-evolution operator
%
%
\begin{equation}
\hat{\Pi}(t)=\left(\begin{array}{cc}
P^{st}_0 & P^{st}_0 \\
P^{st}_1 & P^{st}_1 \end{array}\right)+
\left(\begin{array}{cc}
P^{st}_1 & -P^{st}_0 \\
-P^{st}_1 & P^{st}_0 \end{array}\right) e^{-\Sigma t} \ \  (t>0),
\end{equation}
where we defined the sum rate $\Sigma=\Sigma_{0+}+\Sigma_{1-}$, and the
steady-state occupation probabilities $P^{st}_0=\Sigma_{1-}/\Sigma$ and
$P^{st}_1=\Sigma_{0+}/\Sigma$. The first term in the time evolution 
matrix (operator) $\Pi(t)$ gives the steady-state solution, and the
second term shows simple exponential relaxation with a single rate
$\Sigma$ towards the steady state.
The charge and tunnel current operators are 
\begin{equation}
\hat{N}=\left(\begin{array}{cc}
0 & 0 \\
0 & 1 \end{array}\right) , \
\hat{I}_L=e \Sigma_{0+} \left(\begin{array}{cc}
0 & 0 \\
1 & 0 \end{array}\right), \ 
\hat{I}_R=e \Sigma_{1-} \left(\begin{array}{cc}
0 & 1 \\
0 & 0 \end{array}\right).
\end{equation}
The steady-state properties and correlation functions are
\begin{eqnarray}
\label{ClassNoise}
&&N^{st} = P_1^{st}, \ 
I^{st}_L =  I^{st}_R = P_0^{st} \Sigma_{0+} = P_1^{st} \Sigma_{1-},
\nonumber\\
&&K_{NN}(\tau) =  P^{st}_0 P^{st}_1 e^{-\Sigma \tau},\ \
S_{NN}(\omega)= 
\frac{2I^{st}}{\omega^2 +\Sigma^2}=
\frac{P_0^{st}\Sigma_{0+}+P_1^{st}\Sigma_{1-}}{\omega^2 +\Sigma^2},
\nonumber\\
&&K_{II}(\tau) = \nonumber\\
&&=e I^{st}
[(\alpha_R^2+\alpha_L^2)\delta(\tau)+
(\alpha_L\Sigma_{0+}-\alpha_R\Sigma_{1-})
(\alpha_R\Sigma_{0+}-\alpha_L\Sigma_{1-})
\frac{e^{-\Sigma \tau}}{\Sigma}],
\nonumber\\
&&S_{II}(\omega)=e I^{st}[(\alpha_R^2+\alpha_L^2)+
2\frac{(\alpha_L\Sigma_{0+}-\alpha_R\Sigma_{1-})
(\alpha_R\Sigma_{0+}-\alpha_L\Sigma_{1-})}{\Sigma^2+\omega^2}] =
\nonumber\\
&&=e I^{st}\frac{\Sigma_{0+}^2+\Sigma^2_{1-}+
\omega^2(\alpha_R^2+\alpha_L^2)}{\Sigma^2+\omega^2}.
\end{eqnarray}
Notice that both charge- and current-noise are proportional
to the steady-state current.

\subsection{Low Frequency Quantum Theory}
\label{LowFreqQuantumNoise}
In the finite but low-frequency
regime, i.e. when still only $\Gamma^L_{0+}(\omega)$
and $\Gamma^R_{1-}(\omega)$ are non-zero, the quantum expressions
simplify to:
\begin{eqnarray}
&&S_{NN}(\omega)=\frac{P_0^{st} 2 \Gamma^L_{0+}(\omega)+
P_1^{st} 2 \Gamma^R_{1-}(\omega)}
{\omega^2+(\Sigma_{0+}(\omega)+\Sigma_{1-}(\omega))^2}, \nonumber\\
&&S_{II}(\omega)=2e^2\left[ 
P_0^{st} \alpha_L^2 \Gamma^L_{0+}+
P_1^{st} \alpha_R^2 \Gamma^R_{1-}\right]+
\frac{\alpha_L\Sigma_{0+}-\alpha_R\Sigma_{1-}}
{\omega^2+(\Sigma_{0+}+\Sigma_{1-})^2}\times\nonumber\\
&&\times\left[eI^{st}(\Sigma_{0+}-\Sigma_{1-})+
2e^2(\Sigma_{0+}+\Sigma_{1-})
(P_1^{st}\alpha_R\Gamma^R_{1-}-P_0^{st}\alpha_L\Gamma^L_{0+})
\right],
\nonumber\\
\end{eqnarray}
where the frequency dependence of $\Sigma_{n\pm}(\omega)$ and
$\Gamma_{n\pm}^{L/R}(\omega)$ has been suppressed for
brevity. At zero temperature, with symmetric junctions 
$C_L=C_R$ and $R_T^L=R_T^R=R_T$, the expressions simplify
further:
\begin{eqnarray}
\label{QuantumCorrections}
S_{NN}(\omega)&=&\frac{1}{e}
\left[I^{st}+\frac{\hbar\omega}{e R_T}\right]
\frac{1}
{(\Sigma_{0+}(0)+\Sigma_{1-}(0))^2+\omega^2}, \nonumber\\
S_{II}(\omega)&=&\frac{e}{2}
\left[I^{st}+\frac{\hbar\omega}{2 e R_T}\right]
\left[1+\frac{(\Sigma_{0+}(0)-\Sigma_{1-}(0))^{2}}
{(\Sigma_{0+}(0)+\Sigma_{1-}(0))^{2}+\omega^2}\right] .
\end{eqnarray}
We find that the classically derived expressions (Eq.~\ref{ClassNoise}),
symmetric in $\pm\omega$, get an asymmetric quantum
correction linear in $\omega$. The correction  is added
to the steady-state current and is proportional to
$\hbar\omega/eR_T$, which  would be the current through
one junction with voltage bias $V=\hbar\omega/e$.

Thus the important difference between the orthodox result and the full 
quantum result is that we get an asymmetry in the noise spectrum between 
the positive and negative frequencies in the full quantum result. This 
becomes important for example in a qubit measurement since it will 
drastically change the occupation of the two levels in the qubit.
In the next sections we will discuss such a qubit, and how the noise from 
the SET will affect the measurement of the qubit.

\section{The Qubit}
\label{qubitsection}
The qubit we consider is made up of the two lowest lying energy levels in a
single Cooper-pair box (SCB) \cite{BouchiatPS}. 
An SCB is a small superconducting island, with charging energy
$E_C^{qb}=e^{2}/2(C_{qb}+C_g^{qb}+C_c)$, coupled to a superconducting
reservoir via a Josephson junction with Josephson energy $E_J$. 
In order to have a good qubit the following inequalities have to
be fulfilled: $\Delta_s > E_C^{qb} \gg E_J \gg k_B T$, where $\Delta_s$
is the superconducting energy gap and $T$ is the temperature.
The low temperature is required to prevent thermal excitations and 
the high superconducting gap is needed to suppress quasiparticle tunneling.
For suitable values of the gate voltage (close to $n_g=1/2$)
the box can be described by the following two-level Hamiltonian
\cite{BouchiatPS,MakhlinRMP}
\begin{equation}
\label{eq:qHreduced}
H^{qb}_{q}=-\frac{4 E_C^{qb}}{2} (1-2n_g)\hat{\sigma}_z-
\frac{E_J}{2}\hat{\sigma}_x
\end{equation}
written in the charge basis $\bra{\uparrow}=\bra{n=0}\equiv(1\ 0), 
\bra{\downarrow}=\bra{n=1}\equiv(0\ 1)$,
where $n$ is the number of extra Cooper-pairs on the island, 
$\hat{\sigma}_{x,z}$ are the Pauli matrices, and $n_g=C_{g} V_g/2e$ is
the number of gate-induced Cooper-pairs.
By changing the gate voltage the eigenstates of the qubit can be 
tuned from being almost pure charge states to a superposition of charge
states. 
The eigenstates of the system written in the charge basis are
\begin{eqnarray}
\label{eq:eigenstates}
\ket{0}&=&\cos(\eta/2)\ket{\uparrow}+\sin(\eta/2)\ket{\downarrow}\nonumber\\
\ket{1}&=&-\sin(\eta/2)\ket{\uparrow}+\cos(\eta/2)\ket{\downarrow},
\end{eqnarray}
where $\eta=\arctan(E_J/4E_{qb}(1-2n_g))$ is the mixing angle.
The energy difference between the two states is
\mbox{$\Delta E=\sqrt{(4E_{qb})^{2}(1-2n_q)^2+E_{J}^{2}}$} and the average
charge of the eigenstates is
\begin{equation}
\label{eq:charge}
Q_{0}= 2 e \langle 0|\downarrow\rangle\langle\downarrow|0\rangle
=2e\sin^2(\eta/2), \ \
Q_{1}= 2 e \langle 1|\downarrow\rangle\langle\downarrow|1\rangle
=2e\cos^2(\eta/2).
\end{equation}

\section{SET Charge Measurement of Qubit}
We assume that the qubit is in the state $c_0(0)\ket{0}+c_1(0)\ket{1}$
before a measurement. A perfect charge measurment, i.e. qubit read-out,
will now give the charge $Q_{0}$ or $Q_{1}$ with probability $|c_{0}|^2$
and $|c_{1}|^2$ respectively. 
The two qubit states correspond to slightly different SET gate
voltages, and therefore to two slightly different sets of transition
rates, and two different steady-state currents $I^{st}_0$ and $I^{st}_1$. 
The current fluctuates so there is a finite measurement time $t_{m}$
needed to separate $I^{st}_0$ from $I^{st}_1$\cite{MakhlinRMP}
\begin{equation}
\label{tmTheory}
t_{m}^{theory}=\frac{(I^{st}_1-I^{st}_0)^2}{8 S_{II}(0)}\approx
\frac{e^2 R_T C_\Sigma}{\Delta Q^2} .
\end{equation}
where $\Delta Q=(Q_1-Q_0)\cdot C_c/C_{qb}$ is the charge difference
seen by the SET, and where for the last estimate have we used a 
symmetric SET biased slightly above the
Coulomb threshold at $n_x=0.25$.
The experimentally determined charge sensitivity $\delta Q$
gives the measurement time according to
\begin{equation}
\label{tmExp}
t_m^{exp}=\left(\frac{2\delta Q}{\Delta Q}\right)^2,
\end{equation}
and by comparing Eq.~(\ref{tmTheory}) and Eq.~(\ref{tmExp})
we may deduce a rough theoretical estimate for the charge sensitivity
$\delta Q^{theory}$.
For the parameters of sample \# 1 and \# 2 listed in 
Table\,\ref{Resultstable} we find 
$\delta Q^{theory}=e\sqrt{R_T C_\Sigma}/2 \sim 1.5\cdot10^{-6} e/\sqrt{Hz}$, 
which further substantiates that the measurements are indeed almost
shot noise limited.

\section{SET Back Action on the Qubit}
When performing a measurement on a qubit, the measurement
necessarily dephases the qubit. However, there can also be
transitions between the two qubit states which destroys the information that
the read-out system tries to measure. This mixing occurs due to
charge fluctuations of the SET island $S_{NN}(\omega)$, which
creates a fluctuating charge on the coupling capacitance $C_c$,
which is equivalent to a fluctuating qubit gate charge 
$n_g \rightarrow n_g+\delta n_g(t)$.
The characteristic time for this mixing process is the time $t_{mix}$.
In the weak coupling limit $C_c \ll \{C_\Sigma,C_{qb}\}$, which
is relevant for the qubit measurement setup, we may
use pertubation theory to evaluate $t_{mix}$.

The fluctuating term in the qubit Hamiltonian, written in the charge
basis is
\begin{equation}
\label{eq:voltagefluct}
\delta H^{qb}_{q}(t)=
\frac{4 E_{qb}}{2}2 \delta n_g(t)\hat{\sigma}_z=
E_I \delta n(t)\hat{\sigma}_z, \ \ E_I=\frac{2 e^2 C_c}{C_{qb} C_\Sigma},
\end{equation}
where $e\ \delta n(t)$ represents the fluctuations of charge on the
SET-island, and where we have omitted the term quadratic in $\delta n_g(t)$.
We also defined an electrostatic SET-qubit interaction energy $E_I$.
In the eigenbasis, the full qubit Hamiltonian now reads
\begin{equation}
\label{eq:10}
\delta H_e(t)=-\frac{\Delta E}{2}\hat{\sigma}_z
+E_I \delta n(t)(\cos(\eta)\hat{\sigma}_z+\sin(\eta)\hat{\sigma}_x).
\end{equation}
The $\cos(\eta)$ term leads to dephasing,
and the $\sin(\eta)$ term to interlevel transitions
(level mixing and relaxation).

%
Weak coupling means $E_I \ll \{E_C,E_C^{qb}\}$ and we may use
standard second order time-dependent pertubation theory\cite{Sakurai} 
to see the effect of the charge fluctuations. 
Assuming that the qubit is in a pure coherent state
$c_0(0)\ket{0}+c_1(0)\ket{1}$ at time $t=0$,
we may express the effect of the fluctuating SET charge
in terms of the time-evolution of the quantities
$P^{qb}_{0}(t)=|c_0(t)|^2, P^{qb}_{1}(t)=|c_1(t)|^2$ and
$\xi(t)=|c_0(t)c_1^*(t)|$. In terms of the qubit density-matrix,
$P^{qb}_{0,1}(t)$ are the diagonal elements, determining the occupation
of respective state, and $\xi(t)$ is the magnitude of the off-diagonal 
elements, describing the quantum coherence between the states.
We find that the qubit occupation probabilities obey
a similar master equation as the two-level SET
(Eq.~(\ref{master_equation_two_level})), where now the
transisition rates are determined by the spectral density
of the charge fluctuations at the frequency corresponding
to the qubit level splitting:
\begin{equation}
\label{qubit_relax}
\Sigma^{qb}_{0+} = \frac{E^2_I \sin^2\eta}{\hbar^2} 
S_{NN}(\Delta E/\hbar), \ \
\Sigma^{qb}_{1-} = \frac{E^2_I \sin^2\eta}{\hbar^2}
S_{NN}(-\Delta E/\hbar),
\end{equation}
where $S_{NN}(\omega)$ is the asymmetric charge (number) noise spectral
density defined in Eq.~(\ref{spectral_densities}).
The qubit therefore relaxes exponentially on the timescale $t_{mix}$,
where
\begin{equation}
t_{mix}^{-1}=\Sigma_{qb}=\Sigma^{qb}_{0+}+\Sigma^{qb}_{1-},
\end{equation}
towards the steady-state $P_0^{qb,st}=\Sigma^{qb}_{1-}/\Sigma^{qb}$,
$P_1^{qb,st}=\Sigma^{qb}_{0+}/\Sigma^{qb}$. 
One may note that the quantum corrections in Eq.~(\ref{QuantumCorrections})
cancel in the expression $S_{NN}(\omega)+S_{NN}(-\omega)$ determining
the mixing time.

Due to the charge fluctuations the quantum coherence decays exponentially
\begin{equation}
\xi(t)=\xi(0) e^{-t/\tau_\varphi},\ \
\tau^{-1}_\varphi=2 \frac{E^2_I \cos^2\eta}{\hbar^2}  S_{NN}(0)+
\frac{\Sigma^{qb}}{2},
\end{equation}
where $\tau_\varphi$ is the timescale for dephasing.

\subsection{Signal to Noise Ratio in Qubit Read Out}
\begin{figure}[t]
\centerline{\epsfxsize=10cm\epsffile{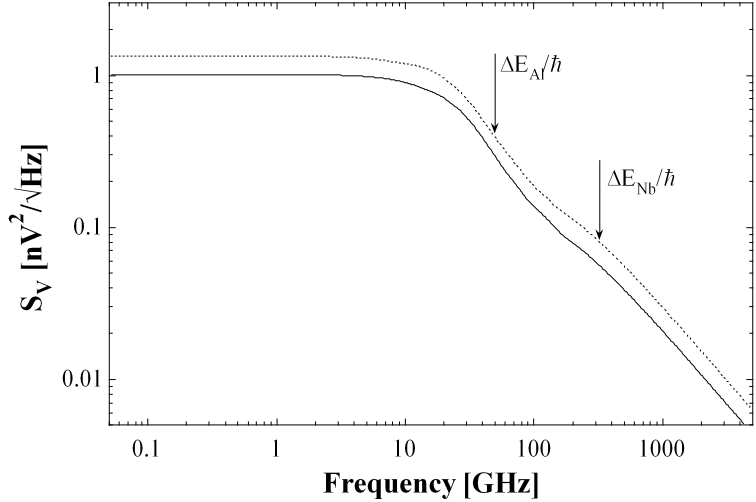}}
\caption{\label{Sv1&2} Calculated $S^{sym}_{VV}(f)$ for samples \#1(full
line) and \#2(dashed line).  The arrows show the energy
separations of the two qubit states for an aluminum and a niobium
qubit, respectively.}
\end{figure}

Now we can use the measured data for samples \#1 and \#2 to
calculate both $t_{m}$ and $t_{mix}$, and thus we can also get the
expected signal-to-noise ratio for a single-shot measurement,
which is simply given by $SNR_{SS}=\sqrt{t_{mix}/t_{m}}$.  The
spectral densities $S_{VV}(f)$ for the two samples at the optimum
charge sensitivity, at a current of 6.7 and 8.0 nA for samples \#1
and \#2, respectively, are displayed in Fig.\,\ref{Sv1&2}.  As can
be seen both through Eq.\,(\ref{qubit_relax}) and in Fig.\,\ref{Sv1&2},
the mixing time increases strongly with increasing $\Delta E$.  In
our case $\Delta E$ is limited by the superconducting energy gap
$\Delta$, which for aluminum films corresponds to about 2.5\,K.
Using niobium as the qubit material would substantially increase
the mixing time. If we assume that $E^{qb}_{C}$ and thus also $\Delta
E$ can be scaled with $\Delta$ and that the coupling $C_c/C_{qb}$
is kept constant, mixing times of several ms can be reached. In
that case, other sources than the SET noise would most probably 
dominate the mixing.
The results for the samples \#1 and \#2 are summarized in
Table\,\ref{Resultstable}, where a coupling $C_c/C_{qb}=0.01$ is assumed.

\begin{table}[b]
\caption{\label{Resultstable}}
\begin{tabular}{c|c|c|c|c|c|c|c}
\hline
SET    & Qubit    & $\delta Q$    & $\Delta E$ & $S_{V}(\omega)$ & $t_{m}$  & $t_{mix}$ & SNR\\
Sample & material & e/$\sqrt{\rm{Hz}}$ & [K]        & [nV$^2$/Hz]     & [$\mu$s] & [$\mu$s]  &    \\
\hline
1 & Al & 6.3 & 2.4  & 0.29  & 0.40 & 8.6  & 4.6\\ 
1 & Nb & 6.3 & 15.5 & 0.056 & 0.40 & 1860 & 68 \\ 
2 & Al & 3.2 & 2.4  & 0.39  & 0.10 & 6.4  & 8.0\\ 
2 & Nb & 3.2 & 15.5 & 0.080 & 0.10 & 1300 & 114\\ 
\hline
\end{tabular}
\end{table}

\subsection{Coulomb Staircase}
\begin{figure}
\centerline{\epsfxsize=8cm\epsffile{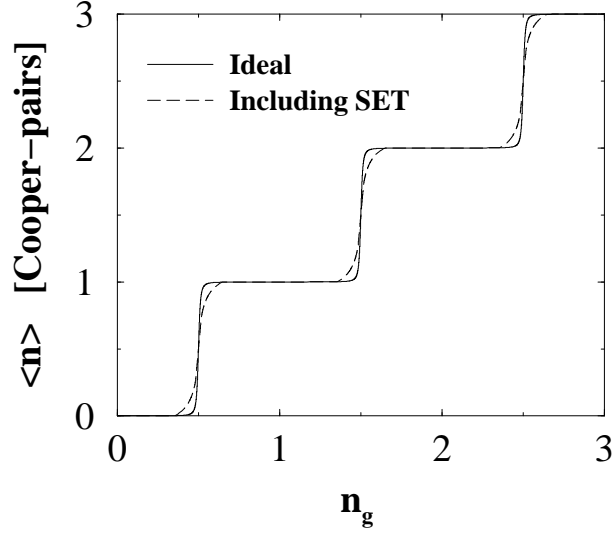}}%
\caption{The calculated Coulomb staircase of an SCB as measured by
the SET, assuming that the SET backaction determine the
steady-state of the qubit.}
\label{CoulombStaircaseFig}
\end{figure}

One may use the SET to measure the so-called Coulomb staircase, 
i.e. the average charge of the qubit as a function of gate voltage.
In an ideal situation with no energy availible from an
external source, at zero temperature, the qubit would follow 
the ground state adiabatically and the charge would increment 
in steps of $2e$ at $n_g=n+0.5$, $n$ integer.
These steps are not perfectly sharp because of the Josephson 
energy mixing the charge states\cite{BouchiatPS}.
Now assuming that the qubit equilibrium is determined by the SET back action
the charge measured should instead be $P_0^{qb,st}Q_0+P_1^{qb,st}Q_1$.
The steps are now rounded further\cite{Nazarov} due to the finite probability
for the qubit to be excited, see Fig.~\ref{CoulombStaircaseFig}.
The quantum asymmetry of the noise-spectrum, indicating the difference
of qubit excitation and relaxation, is vital to recover the
correct Coulomb staircase. Straighforward use of orthodox theory 
would predict an equal population of the qubit states. 

\acknowledgements
One of the authors would like to acknowledge fruitful
discussions with Yuriy Makhlin and Alexander Shnirman.
This work was supported by the Swedish grant
agencies NFR/VR and SSF, the Wallenberg and
G\"oran Gustavsson foundations, and by the SQUBIT
project of the IST-FET programme of the EC.

\newpage

\end{document}